\documentclass[3p,times,procedia]{elsarticle}
\usepackage{nupha_ecrc}

\volume{00}

\firstpage{1}

\journalname{Nuclear Physics A}

\runauth{M. Mackowiak-Pawlowska (the NA61/SHINE Coll.)}

\jid{nupha}

\jnltitlelogo{Nuclear Physics A}

\usepackage{amssymb}
\usepackage[figuresright]{rotating}

\begin{document}

\begin{frontmatter}

%% Title, authors and addresses

%% use the tnoteref command within \title for footnotes;
%% use the tnotetext command for the associated footnote;
%% use the fnref command within \author or \address for footnotes;
%% use the fntext command for the associated footnote;
%% use the corref command within \author for corresponding author footnotes;
%% use the cortext command for the associated footnote;
%% use the ead command for the email address,
%% and the form \ead[url] for the home page:
%%
%% \title{Title\tnoteref{label1}}
%% \tnotetext[label1]{}
%% \author{Name\corref{cor1}\fnref{label2}}
%% \ead{email address}
%% \ead[url]{home page}
%% \fntext[label2]{}
%% \cortext[cor1]{}
%% \address{Address\fnref{label3}}
%% \fntext[label3]{}

\dochead{}
%% Use \dochead if there is an article header, e.g. \dochead{Short communication}
%% \dochead can also be used to include a conference title, if directed by the editors
%% e.g. \dochead{17th International Conference on Dynamical Processes in Excited States of Solids}

\title{NA61/SHINE results on fluctuations and correlations in
p+p and Be+Be interactions at CERN SPS energies}

%% use optional labels to link authors explicitly to addresses:
%% \author[label1,label2]{<author name>}
%% \address[label1]{<address>}
%% \address[label2]{<address>}

\author{Maja Mackowiak-Pawlowska for the NA61/SHINE Collaboration}

\address{Faculty of Physics, Warsaw University of Technology, ul. Koszykowa 75, 00-662 Warszawa, PL}

\begin{abstract}
The aim of the NA61/SHINE strong interaction program is to explore the phase diagram of strongly interacting matter. The main physics goals are the study of the onset of deconfinement and the search for the critical point of strongly interacting matter. These goals are pursued by performing a beam momentum (13$A$ - 158$A$ GeV/c) and system size (p+p, p+Pb, Be+Be, Ar+Sc, Xe+La) scan.

This contribution presents results on transverse momentum and multiplicity fluctuations from the Be+Be and p+p energy scan. Also, results on two-particle correlations in pseudorapidity and azimuthal angle obtained in p+p interactions will be shown. The influence of conservation laws and resonance decays on multiplicity and chemical fluctuations of identified particles in p+p interactions will be discussed. Obtained results will be compared with data from other experiments and with model predictions.
\end{abstract}

\begin{keyword}
%% keywords here, in the form: keyword \sep keyword
proton-proton interactions \sep beryllium-beryllium interactions \sep multiplicity \sep chemical and transverse momentum fluctuation \sep onset of deconfinement \sep two-particle correlations
%% MSC codes here, in the form: \MSC code \sep code
%% or \MSC[2008] code \sep code (2000 is the default)

\end{keyword}

\end{frontmatter}

%%
%% Start line numbering here if you want
%%
% \linenumbers

%% main text
\section{Introduction}
\label{intro}
\vspace{-0.1in}
It is a well established fact that matter exists in different states. For strongly interacting matter at least three states are expected: nuclear matter, hadron gas (HG) and a system of deconfined quarks and gluons (often called the quark-gluon plasma - QGP).

One of the most important goals of high-energy heavy-ion collisions is to establish the phase diagram of strongly interacting matter by finding the possible phase boundaries and critical points. In particular,  we want to produce the quark-gluon plasma and analyze its properties and the transition between QGP and HG. 
This contribution presents experimental results on event-by-event fluctuations of multiplicity, particle type and transverse momentum of charged and identified hadrons produced in inelastic p+p interactions at 20, 31, 40, 80 and 158 GeV/c, and centrality selected $^{7}$Be+$^{9}$Be collisions at 20$A$, 30$A$, 40$A$, 75$A$ and 150$A$ GeV/c. 
The measurements were performed by the multi-purpose NA61/SHINE~\cite{Antoniou:2006mh,Abgrall:2014fa} experiment at the CERN Super Proton Synchrotron (SPS). They are part of the strong interaction program devoted to study of properties of the onset of deconfinement (OD) and search for the critical point (CP) of strongly interacting matter. Within this program a two dimensional scan in collision energy and size of colliding nuclei is in progress. The expected signal of a critical point is a non-monotonic dependence of various fluctuation/correlation measures in such a scan.
\vspace{-0.15in}
\section{Fluctuations}
\vspace{-0.1in}
Evidence for the onset of deconfinement at lower SPS energies suggests the possibility of detecting the CP in heavy ion collisions. A specific property of the CP - the increase in the correlation length - makes fluctuations its basic signal. In order to observe the CP, created matter should freeze-out near its location. So, if at all, the CP signals are expected at energies higher than those of the onset of deconfinement. Indications of enlarged event-by-event fluctuations in multiplicity and transverse momentum as well as an intermittency signal were found by NA49 in Si+Si collisions at 158$A$ GeV~\cite{KG,Anticic:2012xb}. 
	
When comparing fluctuations in systems of different size (volume) one should use quantities which are insensitive to volume and volume fluctuations (unavoidable in A+A collisions). Two families of such quantities were used. First, the intensive quantity scaled variance, $\omega[A] = \frac{\langle A^{2}\rangle-\langle A\rangle^{2}}{\langle A \rangle}$ where $A$ stands for an event quantity. In the Grand Canonical Ensemble it is independent of volume but it depends on volume fluctuations. Second, the strongly intensive quantities $\Sigma[A,B]$ and $\Delta[A,B]$ defined as:
\begin{equation}
	\Delta[A,B] = \frac{1}{C_{\Delta}}\big[ \langle B \rangle\omega_{A} - \langle A \rangle\omega_{B} \big], \quad
	\Sigma[A,B] = \frac{1}{C_{\Sigma}}\Big[ \langle B \rangle\omega_{A} + \langle A \rangle\omega_{B} - 2 \big( \langle AB \rangle - \langle A \rangle \langle B \rangle \big) \Big], 
\end{equation}
where $A$ and $B$ are two different event quantities. There is an important difference between $\Delta[A,B]$ and
$\Sigma[A,B]$. Only the first two moments of A and B distributions are required to calculate $\Delta[A,B]$, whereas $\Sigma[A,B]$ includes the correlation term $\langle AB \rangle - \langle A \rangle\langle B\rangle$. Thus they can be sensitive to various physics effects in different ways. With the proper normalization they are dimensionless and have a common scale~\cite{Gazdzicki:2013ana}. They are 0 for absence of e-b-e fluctuations ($A=const$, $B=const$) and one for the independent particle production model. Here A and B stand for multiplicities of different particle types or multiplicity and sum of transverse momenta. For comparison with the NA49 experiment also the $\Phi$ quantity was used. For fluctuations of $P_{T}=\sum\limits_{k=1}^{N} p_{T_{k}}$ it reads $\Phi_{p_{T}}=\sqrt{\overline{p_{T}}\omega[p_{T}]} \big[ \sqrt{\Sigma[P_{T},N]}-1 \big]$. For particle type fluctuations of two particle types $N_i$ and $N_j$ it is defined as $\Phi_{ij}=\frac{\sqrt{\langle N_{i}\rangle\langle N_{j}\rangle}}{\langle N_{i}+N_{j}\rangle}\cdot[\sqrt{\Sigma[i,j]}-1]$.   

\begin{figure}
\includegraphics[width=0.45\textwidth]{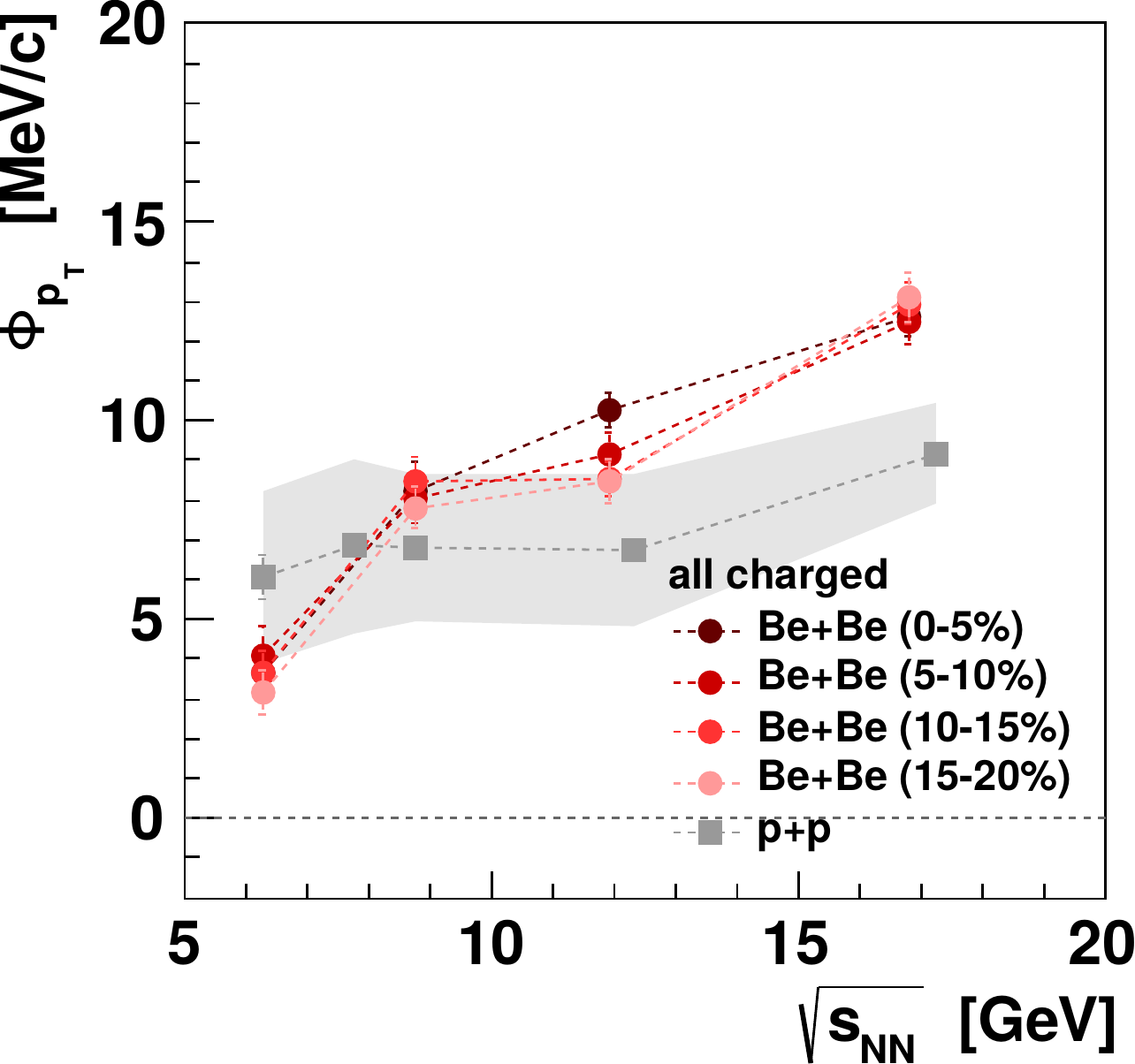}
\includegraphics[width=0.45\textwidth]{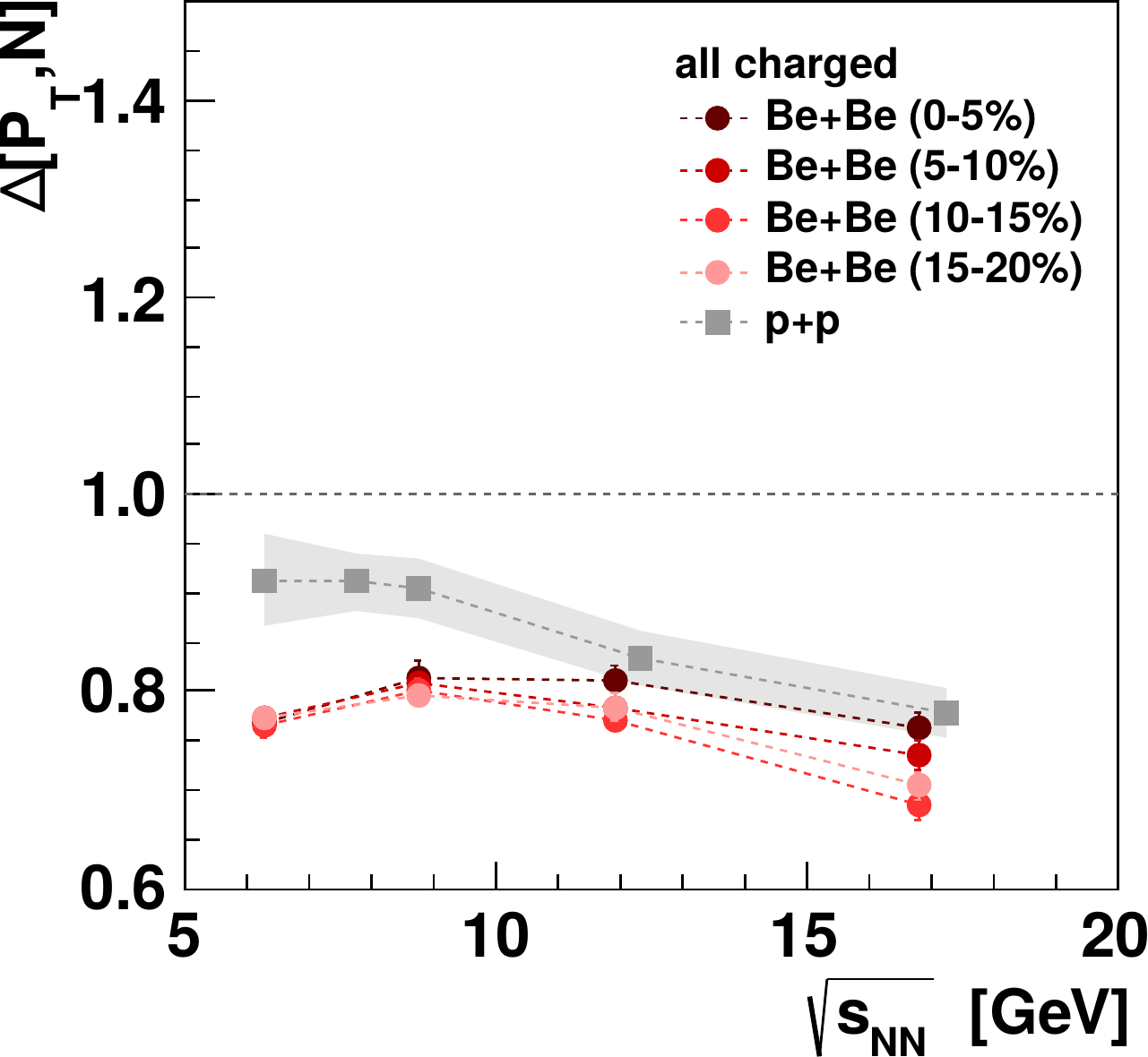}
\caption{Preliminary NA61/SHINE results on the energy dependence of $\Sigma[P_{T},N]$ (left) and $\Delta[P_{T},N]$ (right) in p+p and $^{7}$Be+$^{9}$Be interactions.}
\label{siq}
\end{figure}

Enhanced fluctuations of multiplicity and transverse momentum were suggested as a possible signature of the CP of strongly interacting matter~\cite{Stephanov:1999zu}. Transverse momentum fluctuations were measured by the NA49 experiment using the $\Phi_{p_{T}}$ quantity~\cite{phiref}. For system size dependence at the top SPS energy NA49 finds a maximum of $\Phi_{p_{T}}$ for C+C and Si+Si collisions~\cite{KG}. The energy dependence of $\Phi_{P_{T}}$ and $\Delta[P_{T},N]$ in p+p and $^{7}$Be+$^{9}$Be interactions is shown in Fig.~\ref{siq}. The values in $^{7}$Be+$^{9}$Be interactions exhibit no centrality dependence  and are close to p+p results. No structures which could be connected to the CP or OD in p+p and $^{7}$Be+$^{9}$Be collisions are visible. Bose-Einstein statistics and $\langle p_{T} \rangle$, $N$ correlations probably introduce differences between $\Sigma[P_{T},N]$ and $\Delta[P_{T},N]$. Details on this analysis can be found in Ref.~\cite{::2015jna}.

\begin{figure}
\includegraphics[width=0.45\textwidth]{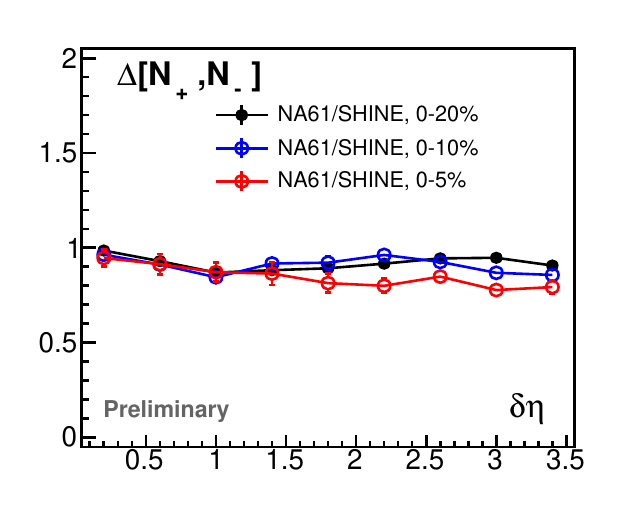}
\includegraphics[width=0.45\textwidth]{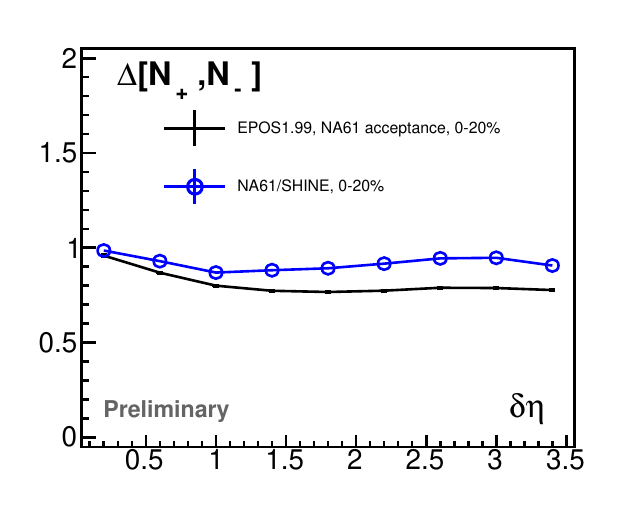}
\caption{Preliminary NA61/SHINE results on $\Delta[N_{+},N_{-}]$ in $^{7}$Be+$^{9}$Be interactions at 150A GeV/c as a function of the size of pseudorapidity window for different size of centrality bins ({\it left}) and with predictions from the EPOS model ({\it right}).}
\label{charge}
\end{figure}

Charge fluctuations may be sensitive to numerous effects including the QGP phase, where the charge carriers are quarks with fractional charge, number of resonances at the chemical freeze-out, conservation laws and long range correlations. Recently, they were studied in $^{7}$Be+$^{9}$Be interactions at 150$A$ GeV/c as a function of the size of pseudorapidity window with center at $\eta\approx4.6$. The size of the $\eta$ window was varied from 0.2 to 3.4. Figure~\ref{charge} shows results on $\Delta[N_{+},N_{-}]$, where $N_{+}$ and $N_{-}$ are multiplicities of positively and negatively charged hadrons. There is no centrality dependence and the data agrees well with the EPOS model~\cite{Werner:2004gh} predictions.
\begin{figure}
\includegraphics[width=0.45\textwidth]{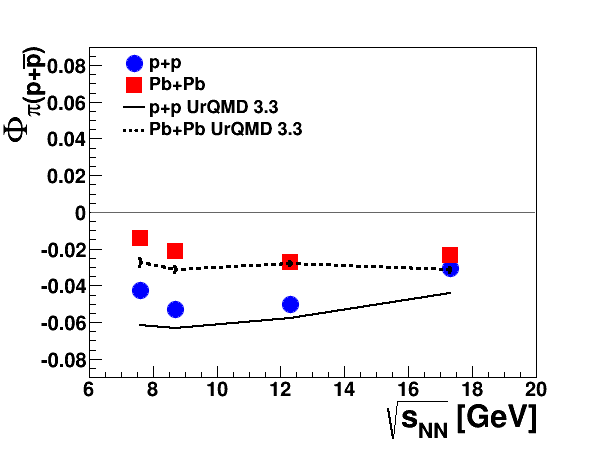}
\includegraphics[width=0.45\textwidth]{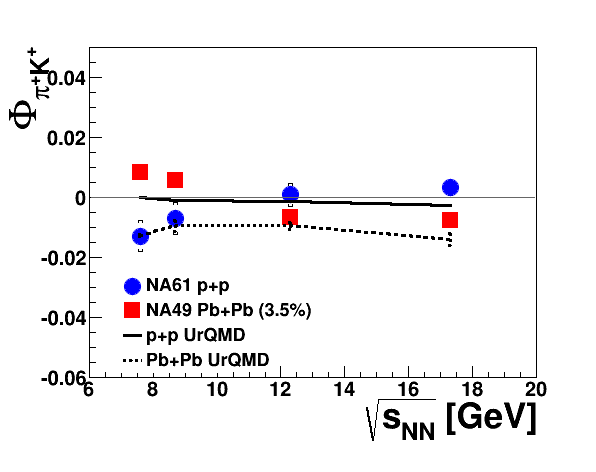}
\caption{The energy dependence of $\Phi_{\pi(p+\bar{p})}$ ({\it left}) and $\Phi_{\pi^{+}K^{+}}$ ({\it right}) in p+p (NA61/SHINE) (preliminary) and central Pb+Pb (NA49) interactions~\cite{Rustamov:2013oza}.}
\label{Figphi}
\end{figure}

Several mechanisms could lead to specific event-by-event particle ratio (chemical) fluctuations. Among them are overheating-supercooling fluctuations due to a first order phase transition with considerable latent heat, or fluctuations due to coexistence of confined and deconfined matter (mixed phase)~\cite{chem}. Comparison of p+p reaction with Pb+Pb interactions from NA49 is shown in Fig.~\ref{Figphi}. The results for $\Phi_{\pi(p+\bar{p})}$ and $\Phi_{\pi^{+}K^{+}}$ in both reactions are similar. $\Phi_{\pi(p+\bar{p})}$ is probably dominated by conservation laws and resonance decays (agreement with UrQMD model supports this interpretation). No structures can be connected with the CP but there is a systematic difference in $\Phi_{\pi^{+}K^{+}}$ possibly connected to OD. Further studies are needed in order to understand the observed discrepancy.
\vspace{-0.15in}
\section{Two particle correlations}
\vspace{-0.1in}
\begin{figure}
\begin{center}
\includegraphics[width=1.8in]{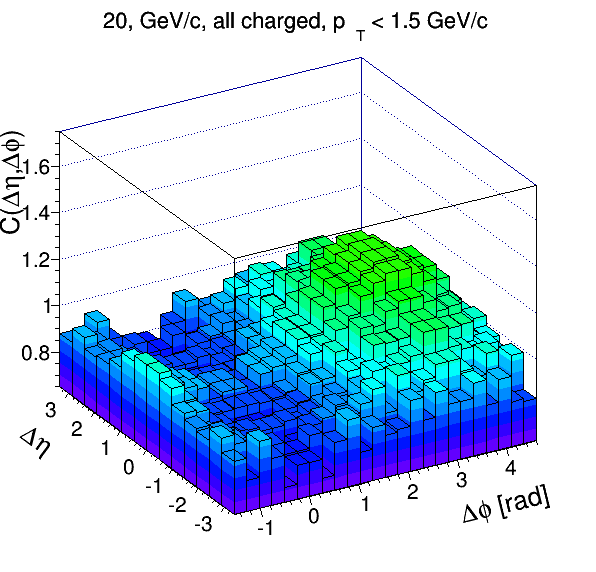}
\includegraphics[width=1.8in]{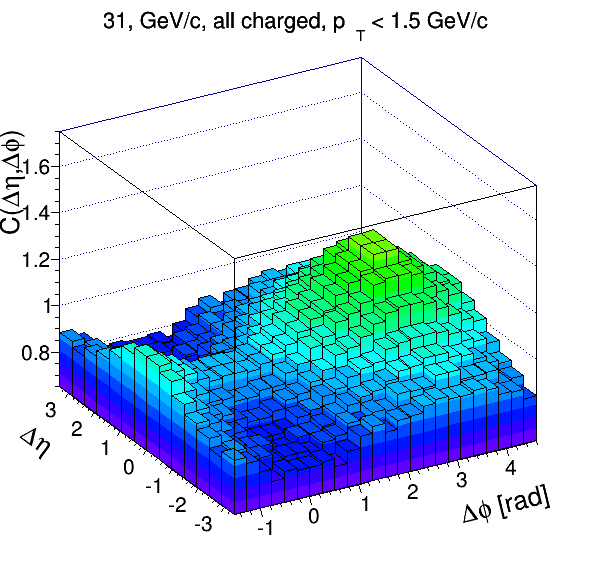}
\includegraphics[width=1.8in]{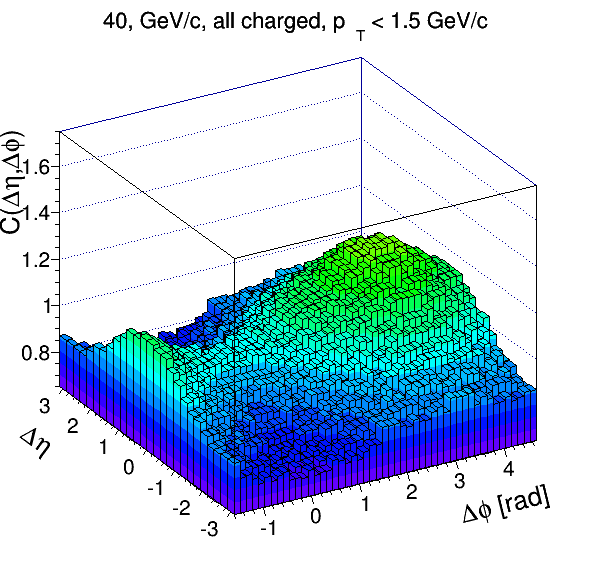}\newline
\includegraphics[width=1.8in]{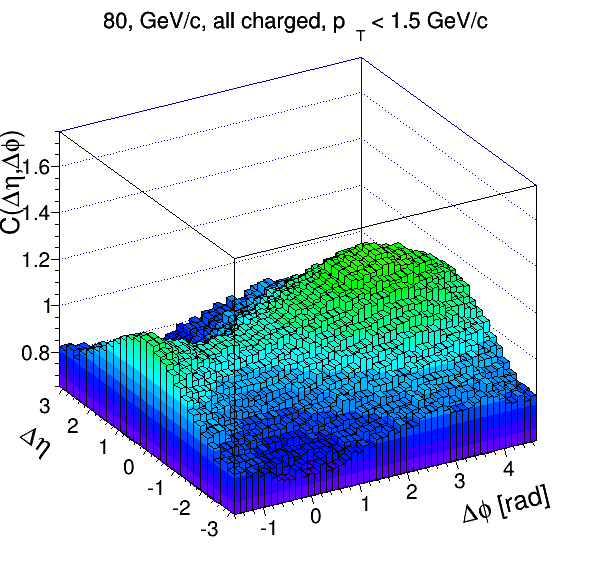}
\includegraphics[width=1.8in]{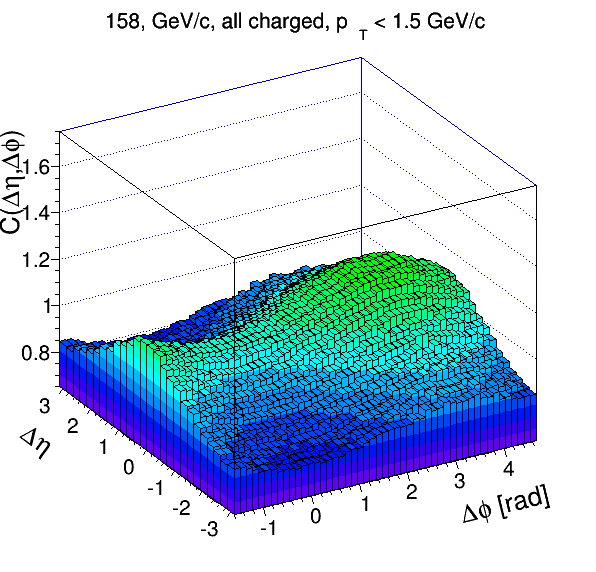}
\end{center}
\caption{Preliminary NA61/SHINE results on correlation function $C(\Delta\eta\Delta\phi)$ in inelastic p+p interactions for all charged particle pairs. The correlation function is mirrored around $(\Delta\eta,\Delta\phi) = (0,0)$.}
\label{deltaphi}
\end{figure}
Two-particle correlations in $\Delta\eta\Delta\phi$ were studied extensively at RHIC and LHC. They may allow to disentangle different sources of correlations: jets, flow, resonance decays, quantum statistics effects, conservation laws, etc. Correlations are calculated as a function of the difference in pseudo-rapidity ($\eta$) and azimuthal angle ($\phi$) between two particles in the same event. The correlation function $C(\Delta\eta,\Delta\phi)$ is obtained as the ratio of distributions of pairs from data and mixed events of the same multiplicity, respectively. Details can be found in Ref.~\cite{Bartek}. The energy dependence of the correlation functions from p+p reactions for all charged pair combinations is presented in Fig.~\ref{deltaphi}.
Two structures can be seen in the plots:
\begin{itemize}
	\item A maximum at $(\Delta\eta,\Delta\phi)=(0,\pi)$,  probably a result of resonance decays and momentum conservation. It is strongest for unlike-sign pairs and significantly weaker for same charge pairs
	\item A weak enhancement at $(\Delta\eta,\Delta\phi)=(0,0)$,  likely due to Coulomb interactions (unlike-sign pairs) and quantum statistics (same charge pairs).
\end{itemize}
\vspace{-0.2in}
\section{Acknowledgements}
\vspace{-0.1in}
This work was supported by the the National Science Centre, Poland grant 2012/04/M/ST2/00816.
%% The Appendices part is started with the command \appendix;
%% appendix sections are then done as normal sections
%% \appendix

%% \section{}
%% \label{}

%% References
%%
%% Following citation commands can be used in the body text:
%% Usage of \cite is as follows:
%%   \cite{key}         ==>>  [#]
%%   \cite[chap. 2]{key} ==>> [#, chap. 2]
%%

%% References with BibTeX database:
\vspace{-0.2in}
\bibliographystyle{elsarticle-num}
%\bibliography{na61References}

%% Authors are advised to use a BibTeX database file for their reference list.
%% The provided style file elsarticle-num.bst formats references in the required Procedia style

%% For references without a BibTeX database:

\end{document}